\documentclass[jkps,showpacs,showkeys,twocolumn]{revtex4} 
\usepackage[pdftex]{graphicx}
\usepackage{amsmath,amsthm,amssymb}
\usepackage{graphicx}
\usepackage{bm}
\usepackage{soul}
\usepackage{verbatim}
\usepackage[usenames,dvipsnames]{xcolor}
\begin{document}


\title[]{Sensing and vetoing loud transient noises for the gravitational-wave detection}
\author{Pil-Jong \surname{Jung$^1$}} 
\author{Keun-Young \surname{Kim$^1$}}
\author{Young-Min \surname{Kim$^3$}}
\author{John J. \surname{Oh$^2$}}
\email{johnoh@nims.re.kr}
\author{Sang Hoon \surname{Oh$^{2}$}}
\author{Edwin J. \surname{Son$^2$}}
\affiliation{$^1$ School of Physics and Chemistry, Gwangju Institute of Science and Technology, Gwangju 61004, South Korea}
\affiliation{$^2$ Division of Basic Researches for Industrial Mathematics, National Institute for Mathematical Sciences, Daejeon 34047, South Korea}
\affiliation{$^3$ School of Natural Science, Ulsan National Institute of Science and Technology, Ulsan 44919, South Korea}

\date{\today}


\begin{abstract}
Since the first detection of gravitational-wave (GW), GW150914, September 14th 2015, the multi-messenger astronomy added a new way of observing the Universe together with electromagnetic (EM) waves and neutrinos. After two years, GW together with its EM counterpart from binary neutron stars, GW170817 and GRB170817A, has been observed. The detection of GWs opened a new window of astronomy/astrophysics and will be an important messenger to understand the Universe. In this article, we briefly review the gravitational-wave and the astrophysical sources and introduce the basic principle of the laser interferometer as a gravitational-wave detector and its noise sources to understand how the gravitational-waves are detected in the laser interferometer. Finally, we summarize the search algorithms currently used in the gravitational-wave observatories and the detector characterization algorithms used to suppress noises and to monitor data quality in order to improve the reach of the astrophysical searches.
\end{abstract}

\pacs{04.30.-w, 04.80.Nn, 07.05.Kf}

\keywords{gravitational-wave, data analysis, detector characterization}

\maketitle

\section{INTRODUCTION}

The detection of gravitational-wave by Advanced LIGO (aLIGO) opened a new window for exploring the mystery of the Universe\cite{i8}. Just two years after the first detection, we became to know the origin of the heavy elements spread in the Universe by observing gravitational-wave and electromagnetic signals coming from the neutron star merger, together with the collaboration between gravitational-wave by aLIGO, Advanced Virgo (AdV), and electromagnetic counterparts observation communities\cite{i3,Abbott:2017wuw,i1,i2,Coulter2017}. 
The ground-based laser interferometer detectors such as LIGO\cite{i4}, Virgo\cite{i5}, GEO 600\cite{i6}, and KAGRA\cite{i7} will be soon operating cooperatively and the new tools for understanding the Universe will be even more exquisite.
%
%

Since the gravitational changes around us is so tiny, the detection of gravitational-waves is very hard and it is only expected to be observed from the energetic events in the Universe. However, apart from the signal coming from the Universe, there are a number of noise sources affecting the gravitational-wave detector that should be suppressed in an appropriate way. The successful detection can be made only if we isolate those noise sources from the real gravitational-wave signals. 

The matched filter used in gravitational-wave search is the most powerful technique to distinguish the gravitational-wave signals from the noises, since we know the analytic waveforms of the gravitational-waves from the motion of binary compact stars. However, its efficiency is challenged when many harmful noises originated from instrumental anomalies of the detector and environmental disturbances nearby detector increase. 
Those noises have high signal-to-noise ratio (SNR) and mimic even the short duration waveforms of high mass binary such as binary black holes or black hole-neutron star binary. This leads to a high false alarm rate of the search result, then the detection meets significant obstacles.
In addition, event though coincidence or coherence in a global network of multiple detectors is used for the unmodeled searches of transient signals coming from the core-collapse supernovae, the high occurrence of transient noises makes unexpected coincident signals among the multiple detectors and it lowers the significance of the search result. 
This is why the detector characterization and the data quality check are necessary and so important. In brief, the investigation of noises that have nothing to do with the real signals can improve the data quality and detection efficiency. 

In this review article, we present the key ingredients in the detection of gravitational-waves, e.g.\ the detector, target sources, and data analysis algorithms for the search and the data quality. We briefly review the sources generating gravitational-waves in Sec.~\ref{sec:2} The basic principles and the noise sources in the laser interferometer detector are summarized in Sec.~\ref{sec:3} The search algorithms in data analysis are classified into two types according to target sources in Sec.~\ref{sec:4} One is the modeled search based on the matched filter, and the other is unmodeled search using wavelets. 
The overview of detector characterization is covered in Sec.~\ref{sec:5}, which focuses on sensing transient noises and vetoing them without removing a potential gravitational wave signal.
Finally, some summary and prospect are given in Sec.~\ref{sec:6}


\section{Gravitational Waves} \label{sec:2}

In the general theory of relativity, the gravitation is described by the geometrical spacetime curved by matter distribution\cite{a1}, which is described by a field equation built upon the general covariance, Einstein field equation\cite{a2,a3,a4,a5,a6}
\begin{equation}
\label{eq:einstein}
{G}_{\mu \nu}=\frac{8\pi G_{N}}{c^2} {T}_{\mu \nu},
\end{equation}
where $G_{N}$ is a Newton's constant and $c$ is a speed of light. Hereafter we set $G_{N}=c=1$.
The gravitational-waves can be derived from the metric fluctuations on the Minkowski spacetime,
${ g }_{ \mu \nu  }={ \eta  }_{ \mu \nu  }+{ h }_{ \mu \nu  }$ with $|{ h }_{ \mu \nu  }|\ll 1$.

Plugging the linearized metric into the Einstein equation~\eqref{eq:einstein} and introducing the trace-reversed metric $\bar{h}_{\mu\nu} = h_{\mu\nu} - (1/2) h$, we obtain a wave equation in the Lorentz gauge
\begin{equation}
\label{eq:lineq}
\Box { \bar{h} }_{ \mu \nu  }=-16\pi { T }_{ \mu \nu  },
\end{equation}
which is nothing but a wave equation describing the propagation of the gravitational-waves\cite{a7}.


Omitting the energy-momentum tensor in Eq.~\eqref{eq:lineq}, the vacuum wave equation is simply written as
\begin{equation}
\label{eq:vacuumeq}
\Box { \bar{h} }_{ \mu \nu  }=0,
\end{equation}
which describes the propagation of the gravitational-waves.
The solution to the vacuum wave equation~\eqref{eq:vacuumeq} is a monochromatic plane wave
\begin{equation}
\label{eq:planewave}
{ \bar{h} }_{ \mu \nu  }={ A }_{ \mu \nu  }\exp(i{ k }^{ \alpha  }{ x }_{ \alpha  }),
\end{equation}
where ${ A }_{ \mu \nu  }$ is the amplitude tensor of the gravitational-wave and ${ k }_{ \alpha  }=(\omega ,{ k }_{ 1 },{ k }_{ 2 },{ k }_{ 3 })$ is the wave vector. 
The null vector condition of ${ k }^{ \alpha }{ k }_{ \alpha }=0$ implies that the gravitational-waves are propagating with the speed of light, and the orthogonality between the amplitude tensor and the wave vector ${ A }_{ \mu \nu  }{ k }^{ \nu }=0$ shows that they are transverse waves. 

\begin{figure}[t!]  
\includegraphics[width=6.6cm]{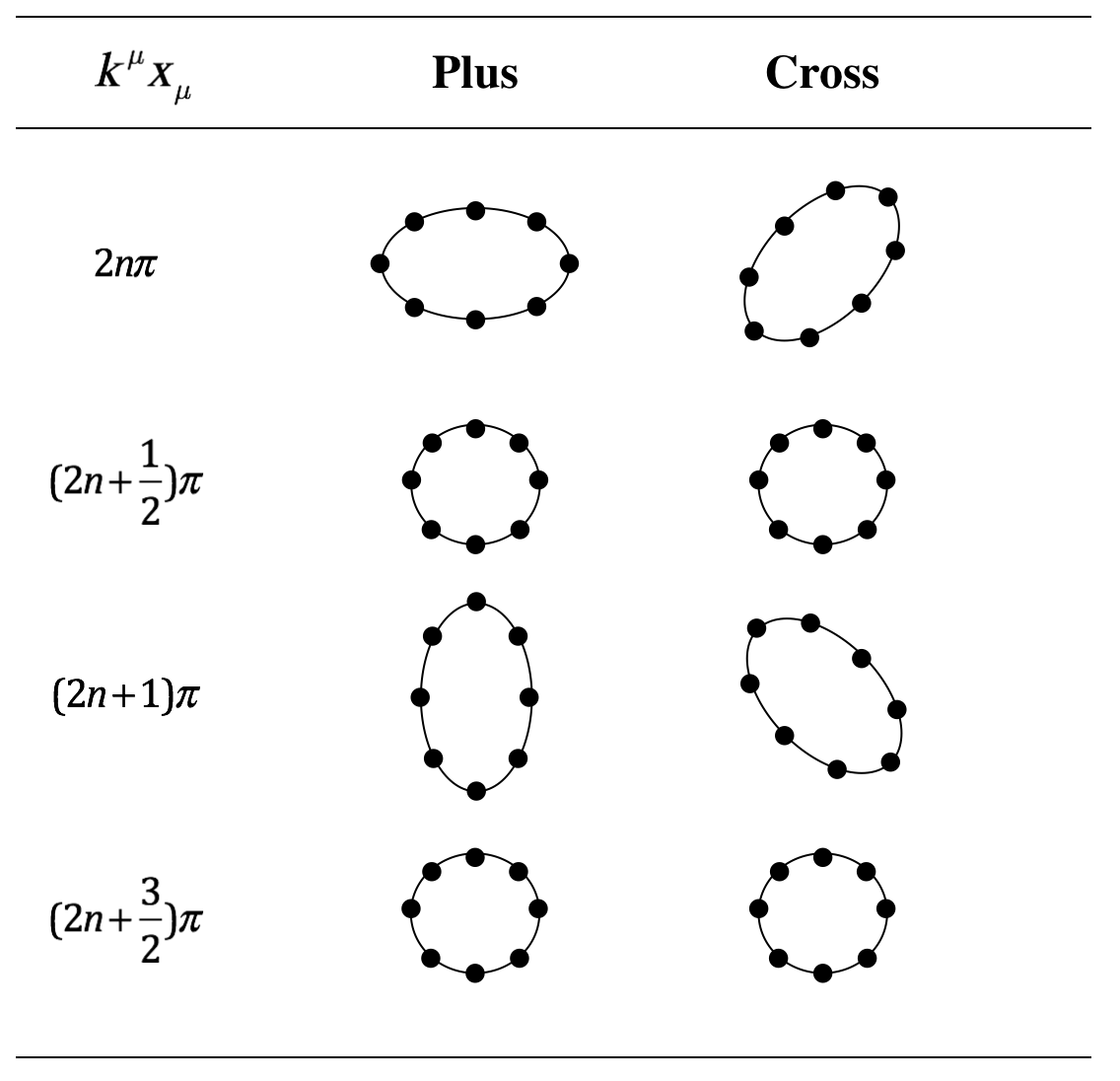}
\caption{Test particles located on the circumference of a circle are oscillating by a gravitational-wave. The drawings show how the circle is changing when the gravitational-wave of plus/cross polarization propagates through the circle.} \label{Fig.1}
\end{figure}

The residual symmetry can be fixed by the transverse-traceless (TT) gauge, and the gravitational-waves are written as
\begin{equation}
{ h }_{ \mu \nu  }^{ TT }=\begin{bmatrix} 0 & 0 & 0 & 0 \\ 0 & { h }_{ + } & { h }_{ \times  } & 0 \\ 0 & { h }_{ \times  } & { -h }_{ + } & 0 \\ 0 & 0 & 0 & 0 \end{bmatrix},
\end{equation}
where $h_+$ and $h_\times$ represent \emph{plus polarization} and \emph{cross polarization}, respectively.
The particle motions with two gravitational-wave polarizations are depicted in Fig.~\ref{Fig.1}.

\subsection{Generation of the Gravitational-Waves}

As the electromagnetic wave is generated by the accelerating charged object, the accelerating massive objects can also generates the gravitational-wave.
The wave equation~(\ref{eq:lineq}) is solved as
\begin{equation}
{ h }_{ \mu \nu  }(x)=-16\pi \int { { d }^{ 4 }x'G(x-x'){ T }_{ \mu \nu  }(x') },
\end{equation}
where $G(x-x')$ is the Green's function and ${ T }_{ \mu \nu  }$ is the stress-energy of a source.
Since this equation depends on boundary conditions, the retarded Green's function can be chosen,
\begin{equation}
G(x-x')=-\frac { 1 }{ 4\pi | \vec { x } -\vec { x' }  | } \delta ({ x }_{ ret }^{ 0 }-{ x' }^{ 0 }),
\end{equation}
where ${x'}^{0}=ct'$, ${x}^{0}_{ret}=c{t}_{ret}$, and ${t}_{ret}=t-|\vec { x } -\vec { x' } |/c$ is the retarded time. Then, the gravitational-wave generated by the source is given by
\begin{equation}
{ h  }_{ \mu \nu  }(x)=4\int { { d }^{ 3 }x'\frac { 1 }{ \left| \vec { x } -\vec { x' }  \right|  } { T }_{ \mu \nu  }(t-\left| \vec { x } -\vec { x' }  \right|/c,\vec { x' } ) }.
\end{equation}

Assuming that the gravitational-wave source is far enough and relatively slow to move, we can reformulate the solution with the \emph{quadrupole moment} ${ Q }_{ ij }$\cite{a8}
\begin{align}
{h }_{ ij }(t,\vec { x } )&=\frac { 2 }{ r } { \ddot { Q }  }_{ ij }(t-r), 
\end{align}
where
\begin{align}
{ Q }_{ ij }&\equiv \int { { d }^{ 3 }x{ T }^{ 00 }(t,\vec { x } ){ x }^{ i }{ x }^{ j } } ,
\end{align}
and the dot represents the derivative with respect to $t$.

Note that the amplitude of the gravitational-wave is proportional to the energy-momentum tensor. Since the observed magnitude of the gravitational-wave is quite small (it decreases as  $1/r$), the gravitational-waves with observable amplitude are limited to violent astronomical and cosmological events such as the collisions of the compact binaries, the supernova explosions, and the cosmic inflation, etc. These gravitational-wave sources will be covered in the next section.

\subsection{The Sources of the Gravitational-Waves}


\begin{table}[tbp] 
\centering
\caption{Some types of the gravitational-wave sources  with their frequency domains}
\label{table}
\begin{tabular}{lll}
\hline
Type & Source & Frequency  \\ \hline 
Transient & compact binary coalescence & 10 Hz - few kHz  \\
	      & supernova explosion	& few kHz \\
Persistent & compact binary inspiral & few mHz  \\
		   & rotating neutron stars & few mHz \\ 
           & primitive background radiation & $10^{-18}$ - $10^{-15}$ Hz  \\ 
\hline
\end{tabular}
\end{table}

The gravitational-waves can be classified into two types; \textit{transient waves} and \textit{persistent waves} (see Table~\ref{table}.). Typical transient gravitational-wave sources are the merger of binary objects, the supernova explosions, and the core-collapse of massive stars. 

The compact binary system emits the gravitational-waves by losing the orbital energy of the two celestial bodies, and then radiates short and strong signal during the merging process\cite{a11,a12,a13}. The waveform of this kind of gravitational-wave signal is called \emph{the chirp signal}, since the amplitude and the frequency of the waveform grow exponentially from inspiral phase to merging phase in time. 
The target detector sensitivity of aLIGO is 40-80 Mpc at early stage, 80-120 Mpc at mid-stage, and 120-170 Mpc for late stage of observation for binary neutron star (BNS) inspirals while 415-775 Mpc at early stage, 775-1\,110 Mpc at mid-stage, and 1\,110-1\,490 Mpc at late stage for binary black hole (BBH) mergers\cite{Aasi:2013wya}.

The event rate of the coalescence of the BNS is expected to be $1\,540_{-1\,220}^{+3\,200}\ \textrm{Gpc}^{-3} \textrm{yr}^{-1}$ after GW170817\cite{i3}, and that of the BBH is 2-600 $\textrm{Gpc}^{-3} \textrm{yr}^{-1}$ after GW150914\cite{Abbott:2016nhf}. According to various observing scenarios\cite{Aasi:2013wya}, the network observation together with aLIGO, AdV, and KAGRA estimates the BNS detection rate 11-180 per year and the localization accuracy of 62-67 \% and 87-90 \% for $5\ \textrm{deg}^2$ and $20\ \textrm{deg}^2$, respectively.

The supernova explosion occurs when a white dwarf in a binary system reaches sufficiently high temperature to ignite nuclear fusion by the companion object or when a massive star's core collapses. The GW signals from core collapse SNe can be observed via GW detector netwrok, which is detectable with aLIGO era out to 20 Mpc\cite{Abbott:2016tdt}.
These events are estimated to occur several times a year within the range of 20 Mpc, and the frequency of the gravitational-waves radiation is below a few kHz\cite{a9,a10}.


Meanwhile persistent sources are a type of the gravitational-wave emitted continuously by the early inspiral phase of the binary system, the rapidly rotating neutron stars, primordial sources, the cosmological inflation, and so on. The astrophysical origin of binary inspiral sources consists mainly of the white dwarfs, neutron stars, and black holes. Since these stars in the inspiral phase emit low-frequency gravitational-waves around a few mHz, direct detection is difficult before the merge. However, since the orbital energy is slowly lost over a long period, indirect detection of the gravitational-waves is possible by observing the change rate of the binary orbit\cite{a14}.

Rotating neutron stars, called pulsars, emit the gravitational waves around 200 Hz frequency by the effects of heavenly pulses which induced by strong magnetic fields and internal instabilities, seismic activity or acceleration\cite{a15,a16,a17,a18}.
The gravitational-waves emitted by stochastic sources are a type of primordial gravitational-wave presumed to have originated in the primitive universe and the background noise that does not interact with the matter after the Planck era. The frequency of this background radiation is about ${10}^{-18}$-${10}^{-15}$ Hz\cite{a19,a20,a21,a22,a23}.


\section{Gravitational-wave detector} \label{sec:3}

The direct detection of gravitational-wave has been tried by J. Weber since 1960s using his resonant bar detector\cite{b1,b2}. Despite of the false detection reports, his pioneering work has lead to long-lasting efforts all over the world to detect the gravitational-wave from the Universe. 

This section introduces the basic idea of laser interferometer as a gravitational-wave detector and various noise sources from instrumental and/or environmental origins that may be harmful against the successful detection of gravitational-waves\cite{b3,b4}.
Moreover, this section covers the enhanced configurations and the noise sources of aLIGO and AdV.

\subsection{Laser Interferometric Detector}

The basic idea of laser interferometric gravitational-wave detector is as follows: First, the incident laser light is split into two arms on the beam splitter. The light reflected from the mirrors located at the end of both arms with the exactly same arm length returns to the beam splitter, and the combined light enters the photon detector. Ideally, if the two test masses at both ends are not affected by the gravitational-waves at all, then the light combined in the beam splitter will has complete destructive interference and will not be measured in the receiver since they travel along the same distance. On the other hand, if the length of both arms changes by the effect of the gravitational-wave polarization, the interference patterns between destructive and constructive interferences will be observed repeatedly\cite{b5}. 

\begin{figure}[t!]  
\includegraphics[width=6.6cm]{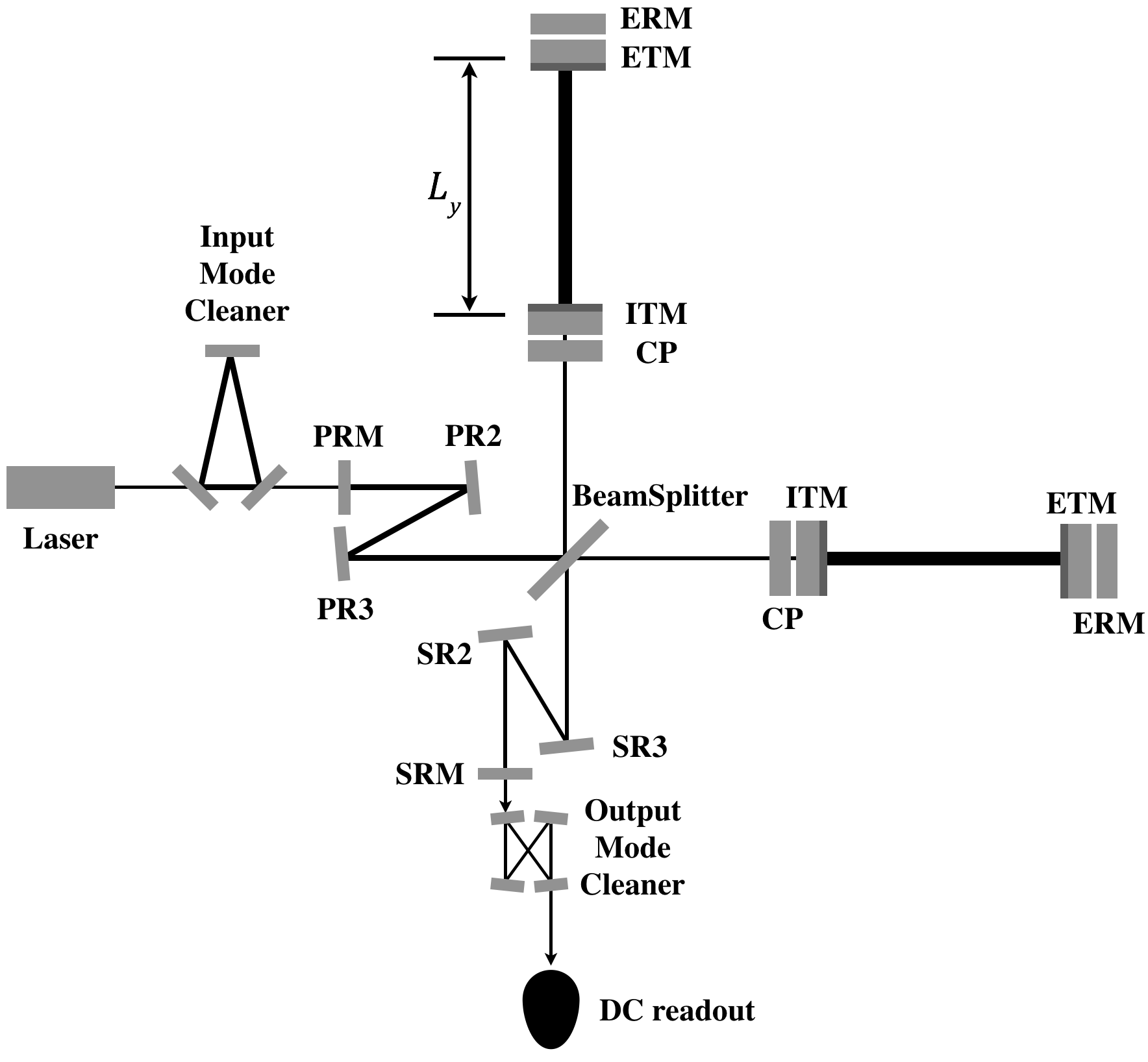}
\caption{The layout of an advanced laser interferometer detector with Fabry-Perot, signal, and power recycling cavities~\cite{TheLIGOScientific:2014jea}} \label{Fig.2}
\end{figure}

Interferometers have constraints to detect the gravitational-waves related to the length of the arms. The phase difference of the Michelson interferometer $\Delta\phi_{mich} \equiv 2 \Delta \phi_{x}(t)$ can be maximized when $\omega L/c = \pi/2$, which leads to
\begin{equation}
\label{eq:arm:length}
L\simeq 750\ \text{km}\left( \frac { 100\ \text{Hz} }{ f }  \right).
\end{equation}

To detect the gravitational-waves nearby 100 Hz, the construction of an interferometer with an arm length of about 750 km should be considered. However, building such huge facilities on the ground is challenging because of the structural and financial problems, so it is necessary to devise a capable method detecting the gravitational-waves with relatively shorter arm length. The Fabry-Perot cavities are designed to satisfy this needs\cite{b6, b7}.

The Fabry-Perot cavities are as simple as folding the arms of an interferometer. The key is to make the effect of using light multi-pass in the arms to create the impact of increasing the effective arm length of the interferometer. So, it has the same result as extending to around 230 km near 100 Hz, which is almost equivalent to the targeted wavelength of the gravitational-wave. In addition to the extending the effective arm length, Fabry-Perot cavity amplifies the laser power inside the cavity, which contains a great amount of photons enhancing the detector sensitivity remarkably. Moreoer, the power recycling cavity is adapted to increase the effective laser power while the signal recycling cavity is added to improve the frequency response\cite{TheLIGOScientific:2014jea}.

The basic idea of observing gravitational-waves using a Michelson interferometer is quite simple, but it is necessary to meet very complex requirements with the high sensitivity required for observations. The laser beam must be focused on the mirror very precisely and must also have the correct wavelength and constant intensity. Multiple reflected beams in the Fabry-Perot cavities must be correctly incident in the beam splitter. In addition, mirrors are required with extremely high level of coating and polishing to prohibit unwanted thermal instabilities caused by scattered beam lights in the cavities\cite{i7,b6,b8}.

\subsection{The Advanced Laser Interferometers}
The initial LIGO (iLIGO) has been operated during 2002-2010 with six science runs. During this period, the iLIGO achieved its designed strain sensitivity with $2\times 10^{-23}/\sqrt{\text{Hz}}$ at 200 Hz. The main noise sources are quantum and suspension thermal noises at mid and high frequencies as well as seismic noise at low frequency. After the initial operation until 2010, the detector started to upgrade with the 10-times better performance design, which is expected to enlarge the observational volume with 1\,000 times. This project is called {\it Advanced LIGO (aLIGO)}. Comparing to the iLIGO, aLIGO has many newly-installed detector configurations with high technologies. For example, it has enhanced pre-stabilized laser, much heavier mirror with high quality factor, challenging requirement of test-mass mirror coating, and so on. (See the detailed configuration about the advanced laser interferometric detectors in Ref.\cite{TheLIGOScientific:2014jea,TheVirgo:2014hva}). The configuration of aLIGO is shown in the schematic figure in Fig.~\ref{Fig.2}. The observing run of the aLIGO started from September 2015 with the sensitivity of $8\times 10^{-24} /\sqrt{\text{Hz}}$ around 100 Hz.

\begin{figure}[t!]  
\includegraphics[width=6.6cm]{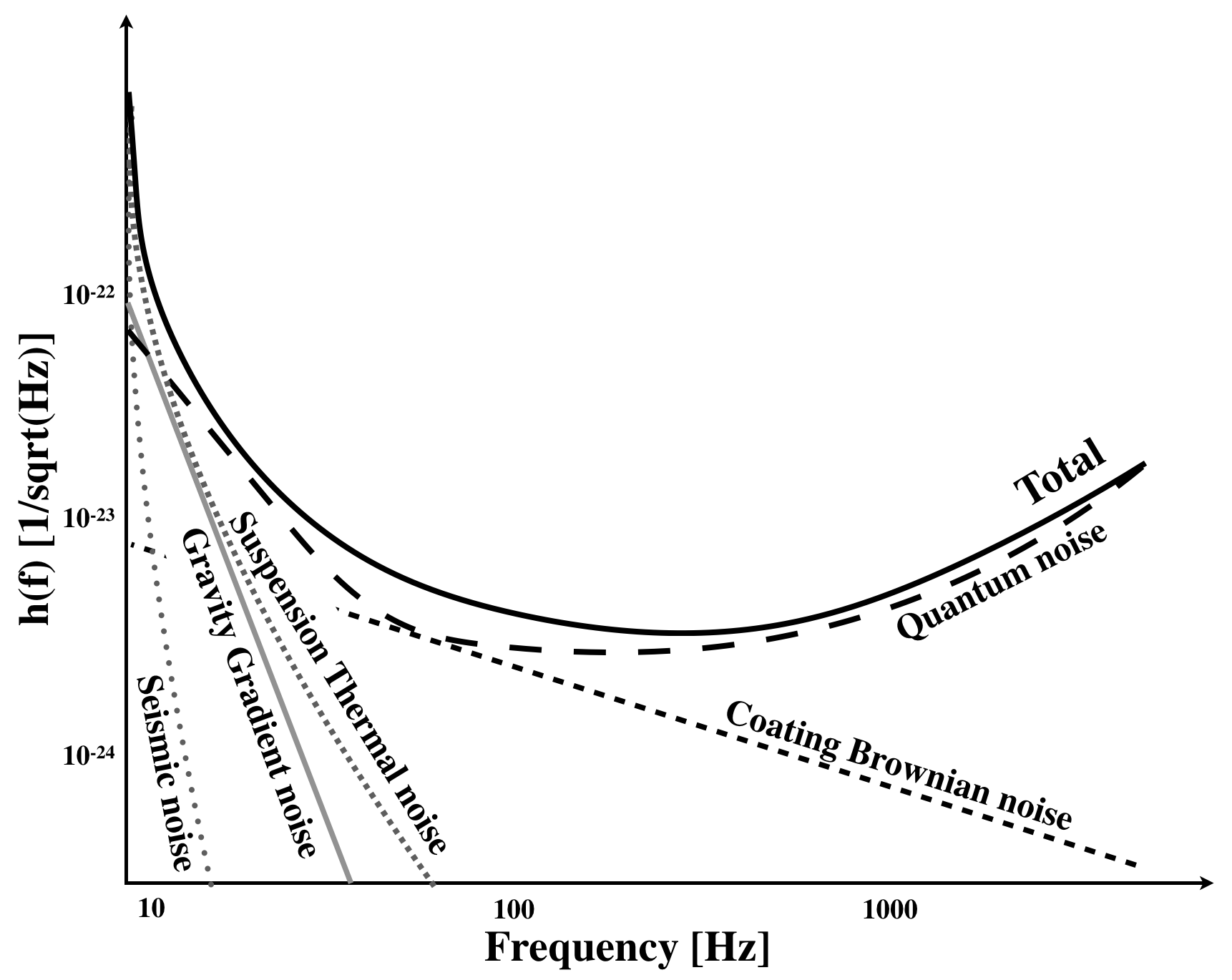}
\caption{A simplified cartoon view of the sensitivity and noise limit curve for aLIGO\cite{TheLIGOScientific:2014jea}} \label{Fig.3}
\end{figure}

To achieve the designed sensitivity, many noise sources should be analyzed in order to reduce them and eliminate the reasons of such noises. Their origins are very diverse but mainly from instrumental and environmental origins: seismic noise, thermal noise, quantum noise (shot noise and quantum radiation pressure noises), etc as shown in Fig.~\ref{Fig.3}.

The sensitivity of aLIGO is governed by three major noise sources -- quantum, seismic, and thermal noises. Below 10 Hz, there exists a seismic variation induced by vibration of ground around the test-masses. Its coupling to the arm length displacement is given by 
\begin{equation}
S_\text{sei}(f) = \frac{2\beta G\rho N_\text{sei}(f)}{(2\pi f)^2},
\end{equation}
where $G_{N}$ is the gravitational constant, $\rho \sim 1\,800\ \text{kg}/\text{m}^3$ is the ground density near mirror, $\beta$ is the geometric factor, and $N_\text{sei}$ is the seismic motion near the test-masses.

Thermal noises can be arisen from various sources of mechanical and/or optical couplings to each components of instruments. One of them is the suspension thermal noise caused by thermal vibrations of the suspension fiber of test-mass mirror\cite{kyama}. Another major contributing source to thermal noise is related to the optical coating of the test-mass mirrors. This is called \emph{coating Brownian noise}. The reduction of this noise is associated with the thickness of the coatings, which also provide the required high reflectivity of the mirrors\cite{coatings}.

The third important noise in the aLIGO is the quantum noise which consists of photon shot noise and quantum radiation pressure noise. This noise is driven by two different types of vacuum field fluctuation through antisymmetric port of the interferometer. The quantum radiation pressure noise is due to the fluctuating radiation pressure force that moves the test-mass mirrors physically. The noise in the arm channel is given by
\begin{equation}
S_\text{rad}(f) = \frac{1.38\times 10^{-17}}{f^2} \left(\frac{P_\text{arm}}{100\ \text{kW}}\right)^{1/2} K_{-}(f) \frac{m}{\sqrt{\text{Hz}}},
\end{equation}
where $P_\text{arm}$ is the power in the arm cavity and $K_{-}(f)$ is the transfer function of negative mode given by
\begin{equation}
K_{-}(f) = \frac{f_{-}}{if+f_{-}}
\end{equation}
with $f_{\pm}$ which is common and differential coupled cavity poles with $f_{+} \sim 0.6$ Hz and $f_{-} \sim 335$-390 Hz, respectively.

Another type of vacuum fluctuation also introduces the photon shot noise. The main laser beam experiences vacuum fluctuation due to optical loss between interferometer and photodiodes. The photon shot noise is given by
\begin{equation}
S_\text{shot}(f) = 2\times 10^{-20} \left(\frac{100\ \text{kW}}{P_\text{arm}\eta}\right)^{1/2} \frac{1}{K_{-}(f)} \frac{m}{\sqrt\text{Hz}},
\end{equation}
where $\eta=0.75$ is the fraction of the power transmitted to the photodiodes.

For other noise sources such as gas noise, charging noise, oscillator noise, etc., see Ref.\cite{Martynov:2016fzi} and references therein. Together with all these noise sources, aLIGO achieved the detector sensitivity of $8\times 10^{-24} /\sqrt\text{Hz}$ during the first observing run and finally succeeded to detect five gravitational waves from the binary black hole mergers and one gravitational-wave from the neutron star binary inspiral until now. With continuous upgrades, the sensitivity will be enhanced much more during aLIGO era. It is also planned to transfer the next generation GW detector era such as A+~\cite{Miller:2014kma}, Voyager~\cite{LSCVoyager}, Cosmic Explorer~\cite{Evans:2016mbw}, and Einstein Telescope~\cite{Punturo:2010zz}.


\section{Data analysis} \label{sec:4}

We have presented the various types of gravitational-wave sources in Sec.~\ref{sec:2} The detection of each type of gravitational-wave requires the corresponding search method. Since the main frequency band of the ground-based laser interferometer detectors is from around 10 Hz to a few kHz, so that the main targets are the compact binary coalescence (CBC)s and the bursts\cite{c0,c1,c2,c3}. 



\subsection{Search for Compact Binary Coalescences}

Since the waveform of the CBCs can be calculated by the post-Newtonian approximation and/or the numerical relativity, the search for the CBC is based on the matched filter with the waveform bank. The LIGO Scientific and the Virgo Collaborations (LVC) have developed PyCBC\cite{Canton:2014ena,Usman:2015kfa,Nitz:2018rgo} and GstLAL\cite{Cannon:2011vi,Privitera:2013xza,Messick:2016aqy} for the CBC search. To infer the physical parameters of the gravitational-wave candidates, Bayesian inference is used with two independent sampling algorithms - Markov Chain Monte Carlo (MCMC) and nested sampling techniques in the parameter estimation process (See Refs.\cite{d17,c14,sluys,vannv,veitch} for the details on the parameter estimation).

PyCBC is a specific pipeline based on the matched filter analysis in frequency domain. To reduce non-Gaussian noises which may have large matched-filter signal-to-noise ratio (SNR), PyCBC additionally calculates chi-squared tests\cite{Canton:2014ena}. The chi-squared statistic computes how different the matched-filter trigger from the waveform and reweight the SNR of each trigger. If triggers from multiple detectors match the same template and appear within a time window of the propagation time plus the uncertainty (5 ms), this coincident trigger is set as an event candidate. When an event candidate is found, its significance is measured by the false alarm rate (FAR), that is, how frequently a background noise trigger occurs with the reweighted SNR equal to or higher than the candidate event.

However, we cannot obtain the pure noise data that is free of the gravitational-wave. To obtain the detection statistic of background noises, the PyCBC introduces the {\it time-slide} technique. The triggers from one detector is arbitrarily shifted in time domain with amount larger than the propagation time of the gravitational-wave to another detector. Then, the new coincident triggers are now free of the gravitational-wave. Since the LIGO and Virgo detectors are sufficiently apart from each other to have no correlated noise, the detection statistic of the background noise is independent of the time-shift.

Another pipeline searching CBC signal, GstLAL performs the matched filter analysis in time domain. Similar to the chi-squared statistic, the GstLAL computes the goodness-of-fit between the measured and expected SNR time series to suppress the large-SNR noises\cite{Cannon:2011vi}. Both the matched-filter SNR and the goodness-of-fit are used as the trigger parameters in the GstLAL. A coincident event is ranked by calculating a likelihood ratio using trigger parameters. To obtain the significance of an event candidate, the GstLAL calculates the likelihood ratio of the background noise using the distribution of triggers that are not coincident in time.

\subsection{Search for Bursts}

The waveform of the gravitational-wave bursts can be calculated only in some specific cases so that we cannot use the matched filter. For this reason, various unmodeled search algorithms have been developed by LVC -- the coherent waveburst (cWB)\cite{c21}, the omicron-LALInference-Burst (oLIB)\cite{c3} algorithms. These methods identify event trigger signals that occur in multiple detectors and reconstruct the waveform of the signal through the maximum likelihood method\cite{c12}. In addition, the BayesWave algorithm with cWB trigger is used\cite{c10,c11}. We briefly introduce these algorithms in this section (for details, see\cite{c23}).

The cWB pipeline is one of the unmodeled search pipelines for the gravitational-wave transients in broad band\cite{c21}. In the time-frequency map of the Wilson-Daubechies-Meyer wavelet transform\cite{Necula:2012zz}, the combined data of two LIGO detectors is used to identify a transient event as a cluster with various frequency resolutions. For each trigger, the data are coherently analyzed to reconstruct the waveforms in the grid of sky locations\cite{c7}. The best waveforms and sky location are selected with a likelihood statistic.

The oLIB\cite{c3} is an unmodeled search pipeline using an event trigger generator called Omicron\cite{d9} which is based on the Q-transform\cite{d10}. In the time-frequency map of the Q-transform, the triggers (excess power) are clustered based on time, central frequency, and quality factor. Then, the coincident triggers are analyzed coherently with a sine-Gaussian wavelet and the Bayesian inference algorithm. Using the time-slide technique, the background triggers are obtained to measure the FAR.

The BayesWave algorithm describes a trigger identified by cWB as a linear combination of sine-Gaussian wavelets\cite{c10}. The wavelets are selected by MCMC for signal and noise models which are characterized by the BayesLine\cite{c11}. The sum of sine-Gaussian wavelets reconstructs the waveform for the signal model. Once the posterior distributions are obtained for each model, the Bayesian inference marginalizes the posteriors to rank the hypotheses.


\section{Detector characterization} \label{sec:5}

The strain data recorded by a ground-based laser interferometer typically contains non-stationary and non-Gaussian noises due to instrument behaviors and environmental issues around the detector. In addition, there exist transient noise signals with various durations due to the causes of instrumental artifacts and environmental influences. The non-stationarity and non-Gaussianity of the data limit search sensitivity by contributing to the search background and lowering the significance of a true astrophysical gravitational wave event. For example, the search for compact binary coalescences based on the matched filtering method might be affected by the loudness of transient noises or transient noises which mimic analytic waveforms. The occurrence rate of transient noises affect coherent searches in a global network of the multiple detectors, such as the unmodeled burst search. 
Moreover, there is potentially correlated noise between detectors such as Schumann resonances\cite{d2}. In this case, we check for lots of physical environmental monitor (PEM) sensors around the time of the correlated noise in addition to performing time slides. PEM sensors are distributed throughout each detector site so that they can monitor potential disturbances to influence a detector. 

In order to convince that GW150914 is not affected by any transient noises as well as any influence from unknown noise sources, data quality (DQ) around the observation time was checked in various ways\cite{d2}. 
Better DQ allows clear search background and higher statistical significance of signal detection, and during the first observing run of the Advanced LIGO, the impact of DQ is shown clearly on GW151226 \cite{o1dq}. 
The summary of the impact of DQ on the CBC search during the first observing run of the Advanced LIGO is addressed in Ref.\cite{o1dq}. 
The searches for long duration continuous waves and gravitational-wave backgrounds are affected by various spectral lines of long duration noise artifacts at a given frequency. The study on indentification and mitigation of narrow spectral artifacts during the first and second observing runs is summaried in Ref.\cite{covas2018}. Continuous wave signals and gravitational-wave background have not been detected yet, but the study on the spectral lines can be used for future observation.

The time when the strain data is too noisy and contains non-Gaussian transient noises is not suitable to be analyzed for searching a gravitational-wave signal originated from an astrophysical source. 
In order to mitigate noise sources and improve the search sensitivity, we need to identify the cause of the problems and sense the abnormal artifacts that occur during the whole operation of the detectors. 
Potential noise sources and/or the phenomena originated from those sources should be recorded and investigated. These instrumental and environmental issues are monitored and recorded by tremendous sensors which are installed in and around the detectors, and the sensors are called auxiliary channels. By using the information gathered from the auxiliary channels, any disturbances and noise sources are investigated, and the results of the study on the characteristics of the noises are used to mitigate the effect of the artificial noises. The summary of all kinds of the characteristics of the noises manifests data quality (DQ) of the detector.
If we identify a noise and fully understand the mechanism of the noise source, we apply an appropriate technique to software and hardware configuration \cite{d2,o1dq,covas2018,d21} or directly use the mitigation method\cite{d21} into data such as {\it gating}\cite{Usman:2015kfa} or {\it noise substraction}\cite{driggers2018}. A best way to mitigate the effect of the noise is to use the characteristics of the noise for tracking down the corresponding noise source, and fix the problem so that it no longer affects $h(t)$ data.

Since fixing directly the problem of the noise is not always possible, there exist remained noise sources and/or the coupling mechanisms between $h(t)$ and the witness channels of the noise which are not fully understood.
The time segments corresponding those noises cannot be removed by a simple way and are marked as DQ vetoes. DQ vetoes are applied depending on a specific search analysis described in the previous section, and the corresponding time is rejected before the search analysis.
DQ vetoes are basically generated by the identification of the noises, the studies of their sources, and the coupling mechanisms between $h(t)$ and the witness channels of the noises. 

DQ vetoes are categorized depending on the severity of the problem or the impact on a search's background. 
Category 1 DQ vetoes are applied to time segments when the detector is not in nominal operation due to a critical issue with a key component of a detector. The data taken during that time should not be analyized for searching a gravitational wave signal. 
Category 2 DQ vetoes are applied after an inital process of data for a specific search. The noise source and its physical coupling mechanism to $h(t)$ in category 2 DQ must be understood, and there should be no possiblity to remove a potential gravitational wave signal. 
Category 3 DQ vetoes are identified to have statistically corelated with $h(t)$, but not completely understood and the further investigation is needed. Category 3 DQ vetoes are used depending on a search pipeline for a cleaner search background. 
When DQ vetoes are applied to a search pipeline, a potential gravitational-wave signals should not be removed and the improvement on the search background distribution should be significantly shown\cite{d21}. 
The safety study of DQ products is briefly described in the later part of this section. 

The studies of data qualities have been conducted in multi-stages of LIGO and Virgo operations and the improvement of the data quality has been achieved. 
The improvement of the data quality during LIGO's 6th Science run (S6) and Virgo's 2nd and 3rd runs (VSR2 and VSR3) are summaried in Ref.\cite{mciver2012}, and \cite{nuttall2015} summarizies the improvement during the early engineering runs right before the starting of Advanced LIGO's first observing run. 
Exmaples of identification and mitigation for specific noises during the first and second observing runs of the Advanced LIGO are also addressed in Ref.\cite{berger2018,nuttall2018,walker2017}. 
In addition, due to a huge amount of auxiliary channel data for noise investigation, machine learning algorithms (MLAs) such as an artificial neural network, random forest, and support vector machine can be adopted to analyze the multidimensional auxiliary channel data, and those have already tested for noise identification\cite{auxmvc}, and it showed comparable performance with a well-known veto method. More advanced techniques of MLA are now applied to noise classification\cite{powell2016,zevin2017}. Especially the {\it Gravity spy}\cite{zevin2017} project is opened to citizen scientist to contribute the noise classification.

In this article, we focus on transient signals, therefore, the rest of the section for detector characterization deals with sensing and vetoing transient signals. 
More detailed techniques and the results of the studies on various noises and the veto effectiveness are well described in many articles \cite{d2,o1dq,d1,d3,d4,d5,d6,d7,d9,d11,d8,d12,d13,dd1,d14,d10,d16,nuttall2015,berger2018,nuttall2018,zevin2017,walker2017,powell2016,mciver2012,covas2018,driggers2018,d21,auxmvc} and the references therein.

\subsection{Sensing : Event Trigger Generator}


In this subsection, we introduce how to identify a transient signal in time series data based on an excess energy search algorithm. 
A short-time segment identified by a transient search algorithm is called \textit{a trigger}, and the well-known event trigger algorithms are \textit{Omicron}\cite{d9,d10,d11} and \textit{Kleine-Welle}\cite{d8,d10}.
\textit{Omicron}\cite{d9,d10,d11} is a transient search algorithm based on Q-transform\cite{d10}, and widely used in LIGO and Virgo collaboration, especially for identifying transient noises in $h(t)$ as well as unexpected disturbances in auxiliary channels. 
In a period of the first observing run of the Advanced LIGO, \textit{Omicron} triggers were used for characterizing the LIGO detectors\cite{d2,o1dq}. 
\textit{Kleine-Welle} (KW) trigger generation is based on the discrete dyadic wavelet transform \cite{d8,d10} and KW triggers have been used since the early stage of LIGO operation. 
We, here, focus on a sensing method with Omicron based on the Q-transfrom.

    
Time series data collected from the detector is analyzed based on segments with minimal time duration for the analysis pipeline.
Each segment is basically calculated by Fourier transformation and then normalized with power spectral density (PSD).  From this procedure, we can obtain triggers from each segment. We will not explain the detailed descriptions here but see Ref. \cite{d10} and references therein for more details.

Basically the trigger is identified by the excess power in the data, which can be parametrized with the central time (${t}_{c}$), the central frequency (${f}_{c}$), the time duration (${\sigma}^{2}_{t}$), and the bandwidth (${\sigma}^{2}_{f}$) defined by
\begin{eqnarray}
&&{ t }_{ c }=\int _{ -\infty  }^{ +\infty  }{ { t\frac { { |h(t)| }^{ 2 } }{ { \left< { h }^{ 2 } \right>  } }  }dt },~~ { f }_{ c }=\int _{ -\infty  }^{ +\infty  }{ { f\frac { { |\tilde { f } (f)| }^{ 2 } }{ { \left< { h }^{ 2 } \right>  } }  }df }, \nonumber\\ && { { \sigma  }^{ 2 } }_{ t }=\int _{ -\infty  }^{ +\infty  }{ { { (t-{ t }_{ c }) }^{ 2 }\frac { { |h(t)| }^{ 2 } }{ { \left< { h }^{ 2 } \right>  } }  }dt },  \nonumber\\ && { { \sigma  }^{ 2 } }_{ f }=\int _{ -\infty  }^{ +\infty  }{ { { (f-{ f }_{ c }) }^{ 2 }\frac { { |\tilde { f } (f)| }^{ 2 } }{ { \left< { h }^{ 2 } \right>  } }  }df },
\end{eqnarray}
where the squared amplitude of detected signal in time and frequency domain is defined as
\begin{equation}
{ \left< { h }^{ 2 } \right>  }=\int _{ -\infty  }^{ +\infty  }{ { |h(t)| }^{ 2 }dt } =\int _{ -\infty  }^{ +\infty  }{ { |\tilde { h } (f)| }^{ 2 }df } . 
\end{equation}
The quality factor $Q$ is defined based on the assumption that duration and bandwidth have the uncertainty relationship ${\sigma}_{t}{\sigma}_{f} \ge  {1}/{4\pi}$,
\begin{equation}
Q=\sqrt { 2 } {f}_{c}/{\sigma}_{f}.
\end{equation}


The time series data corresponding to each segment is projected onto the 3-dimensional parameter space tiles in a finite region of parameter space, $\left[ { t }_{ c }^{\rm min };{ t }_{ c }^{\rm max } \right] \times \left[ { f }_{ c }^{\rm min };{ f }_{ c }^{\rm max } \right] \times \left[ { Q }^{\rm min };{ Q }^{\rm max } \right] $ with a constant quality factor $Q$. 
The density of the tile should be maximized to obtain the fastest analysis speed. However, the number of tiles must be determined with a minimum density to ensure the highest detection efficiency. Since the trade-off between these two conflicting concepts is required, the parameter space tiles are determined so that the fractional energy loss due to mismatch among tiles can be smaller than the threshold, ${\mu}_{\rm max}$.
The mismatch metric of the tiles used in Omicron is given by 
\begin{equation}
\delta { s }^{ 2 }=\frac { 4{ \pi  }^{ 2 }{ f }^{ 2 }_{ c } }{ { Q }^{ 2 } } \delta { t }_{ c }^{ 2 }+\frac { 2+{ Q }^{ 2 } }{ 4{ f }^{ 2 }_{ c } } \delta { f }^{ 2 }_{ c }+\frac { 1 }{ 2{ Q }^{ 2 } } \delta { Q }^{ 2 }.\label{eq:mismatch}
\end{equation}
Note that the non-diagonal term of $\delta \phi \delta Q$ in the metric has been neglected here.

The minimum number of tiles, ${N}_{{t}_{c}} \times {N}_{{f}_{c}} \times {N}_{Q}$ which satisfies the threshold, ${\mu}_{\rm max}$ is obtained from the integration of the mismatch metric, Eq.~\eqref{eq:mismatch} over the three dimensions as follows
\begin{eqnarray}
&&{ N }_{ { t }_{ c } }\ge \frac { { s }_{ { t }_{ c } } }{ 2\sqrt { { \mu  }_{\rm max }/3 }  }, ~~    { s }_{ { t }_{ c } }=\frac { 2\pi { f }_{ c } }{ Q } ({ t }_{ c }^{\rm max }-{ t }_{ c }^{\rm min })~, \nonumber\\&& { N }_{ { f }_{ c } }\ge \frac { { s }_{ { f }_{ c } } }{ 2\sqrt { { \mu  }_{\rm max }/3 }  }~ ,  ~~ { s }_{ { f }_{ c } }=\frac { \sqrt { 2+{ Q }^{ 2 } }  }{ 2 } \ln { ({ f }_{ c }^{\rm max }/{ f }_{ c }^{\rm min }) } ~, \nonumber\\ &&{ N }_{ Q }\ge \frac { { s }_{ Q } }{ 2\sqrt { { \mu  }_{\rm max }/3 }  }~ , ~~{ s }_{ Q }=\frac { 1 }{ \sqrt { 2 }  } \ln { ({ Q }^{\rm max }/{ Q }^{\rm min }) }~, 
\end{eqnarray}
where $N_Q$, $N_{f_c}(Q^q)$, and $N_{t_c}(Q^q,f_c^{ql})$ are the numbers of logarithmically-spaced $Q$ planes, logarithmically-spaced frequency rows in each $Q$ plane, and linearly-spaced time bins, respectively. 
Using the above relations, the resolution of the optimized tile can be determined as follows
\begin{eqnarray}
&&{ Q }^{ q }={ Q }^{ min }{ \left[ \frac { { Q }^{ max } }{ { Q }^{ min } }  \right]  }^{ (0.5+q)/{ N }_{ q } }~~  (0\le  q   < { N }_{ Q }), \nonumber\\
&&{ f }_{ c }^{ ql }={ { f }_{ c } }^{ min }{ \left[ \frac { { { f }_{ c } }^{ max } }{ { { f }_{ c } }^{ min } }  \right]  }^{ (0.5+l)/{ N }_{ { f }_{ c } }({ Q }^{ q }) }~~(0\le  l <{ N }_{ { f }_{ c } }({ Q }^{ q })), \nonumber\\
&&{ t }_{ c }^{ qlm }={ t }_{ c }^{ min }+\frac { (m+0.5)({ t }_{ c }^{ max }-{ t }_{ c }^{ min } )}{ { N }_{ { t }_{ c } }({ Q }^{ q }, { f }_{ c }^{ ql }) }\nonumber\\
&&\qquad\qquad\qquad\qquad\qquad~~(0\le  m <{ N }_{ { t }_{ c } }({ Q }^{ q },{ f }_{ c }^{ ql })). 
\end{eqnarray}


Q-transform is a modified short-time Fourier transform re-defined on the parameter space tiles of a constant quality factor $Q$ with optimized resolutions\cite{dd1,d9}. According to the definition of Q-transform\cite{d9},
the transform coefficient $X$ for a time series $h(t)$ can be obtained by
\begin{eqnarray}
 X({ t }_{ c },{ f }_{ c },Q)&=&\int _{ -\infty  }^{ +\infty  }dt{ h(t)\omega (t-{ t }_{ c },{ f }_{ c },Q)e^{-2i\pi { f }_{ c }t}}  \nonumber\\ &=&\int _{ -\infty  }^{ +\infty  }df{ \tilde { h } (f+{ f }_{ c }){ \tilde { \omega  }  }^{ * }({ f,f }_{ c },Q)e^{2i\pi { f }{ t }_{ c }}}.
 \end{eqnarray}
The window function, $\tilde { \omega  } ({ f,f }_{ c },Q)$ in frequency domain is Gaussian and real, then
\begin{equation}
\tilde { \omega  } ({ f,f }_{ c },Q)={ \tilde { \omega  }  }^{ * }({ f,f }_{ c },Q)={ W }_{ g }{\rm exp}\left( -\frac { { f }^{ 2 } }{ 2{ { f }_{ c } }^{ 2 } }  \right).
\end{equation}
By using normalization condition such as
\begin{equation}
\int _{ -\infty  }^{ +\infty  }{ { |\tilde { \omega  } (f,{ f }_{ c },Q)| }^{ 2 }df } =2,
\end{equation}
then, the normalization factor ${ W }_{ g }$ is given by
\begin{equation}
{ W }_{ g }=\left( \frac { 2^{1/2} }{ \pi^{1/2}} \frac { Q }{ { f }_{ c } } \right)^{1/2}
\end{equation}

\begin{figure}[t!]  
\includegraphics[width=6.6cm]{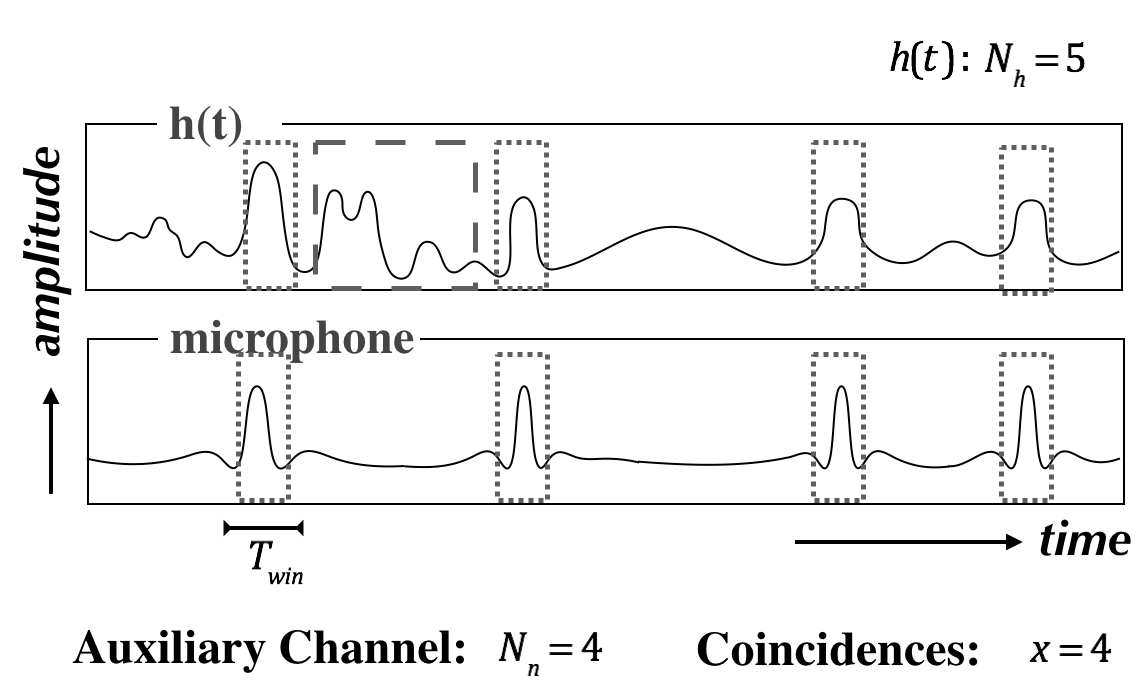}
\caption{Illustration of how to determine the triggers in channels, and the coincidences between the triggers of $h(t)$ and an auxiliary channel\cite{d16}.} \label{Fig.4}
\end{figure}

The list of tiles generated by the Omicron is a set of $(q,l,m)$ tiles with signal-to-noise ratio (SNR), $\rho_{qlm}$ which is defined by 
\begin{equation}
{ \rho_{qlm}  }^{ 2 }=\frac { { |X_h({ t }_{ c }^{ qlm },{ f }_{ c }^{ ql },{ Q }^{ q })| }^{ 2 } }{ \left<| { { X }_{ n}({ t }_{ c }^{ qlm },{ f }_{ c }^{ ql },{ Q }^{ q }) }|^{ 2 } \right> /2 }, \label{eq:Osnr}
\end{equation}
where ${ X }_{ h }$ is the $Q$-transform coefficient of $h(t)$ which is a waveform we want to detect and ${ X }_{ n }$ is the coefficient of $n(t)$ which is noise of a detector. In general, the time series data $x(t)$ is described as $x(t) = h(t) + n(t)$.  $\left<\cdot \right>$ in the denominator of Eq.(\ref{eq:Osnr}) means the expectation value of $Q$ transform energy for noise obtained from multiple measurement. For a stationary stochastic noise, the expectation value can be regarded as the noise power spectral density integrated over the frequency window of tile $(q,l,m)$.
When the SNR of a tile is higher than a given threshold, $\rho_{\rm min}$, which means the tile's energy is higher than the background noise spectrum on given localized parameter sets, we call it a \textit{trigger} (See Fig.~\ref{Fig.4}). The SNR is the standard of the trigger selection in gravitational wave data analysis that varies with specific data analysis pipelines.
In the manner of matched-filtering method, the Omicron triggers can be regarded as the results of applying the matched filters based on sine-Gaussian waveforms in time-frequency planes with various Q-values.

\subsection{Vetoing : Transient Characterization}

In general, if the origins of the many noises detected at the beginning of the observation are recognized or entirely understood, these noises can be precisely excluded from the analysis. However, since it is difficult to comprehend the precise origin even if the triggers are recognized through the sensing process, 
we use statistical methods to figure out which triggers are vetoed before applying search pipeline. 
In LIGO and Virgo collaboration, there are statistical methods utilized to satisfy the requirement. 
Hierarchical veto method (h-veto)\cite{d16}, used percentage veto (UPV) method\cite{d10}, and ordered veto list (OVL) method\cite{d7} are well-known vetoing methods, and the DQ information generated by the methods are provided to commissioners, instrumentalists, and data analysts who want to conduct further investigation or background evaluation through LIGO DQ missions. 
Escpecially, OVL is used for the generation of low-latent DQ information, which is one of the DQ tools used for on-line DQ summary to be delivered when an candidate event appears.
These methods are effective for the study of the background noise reduction in gravitational-wave signals and helps in improving detector performance and data quality. 
Among three vetoing methods, h-veto and UPV are briefly reviewed in the rest of this section.

	\subsubsection{hierarchical-veto method}

\begin{figure}[t!]  
\includegraphics[width=6.6cm]{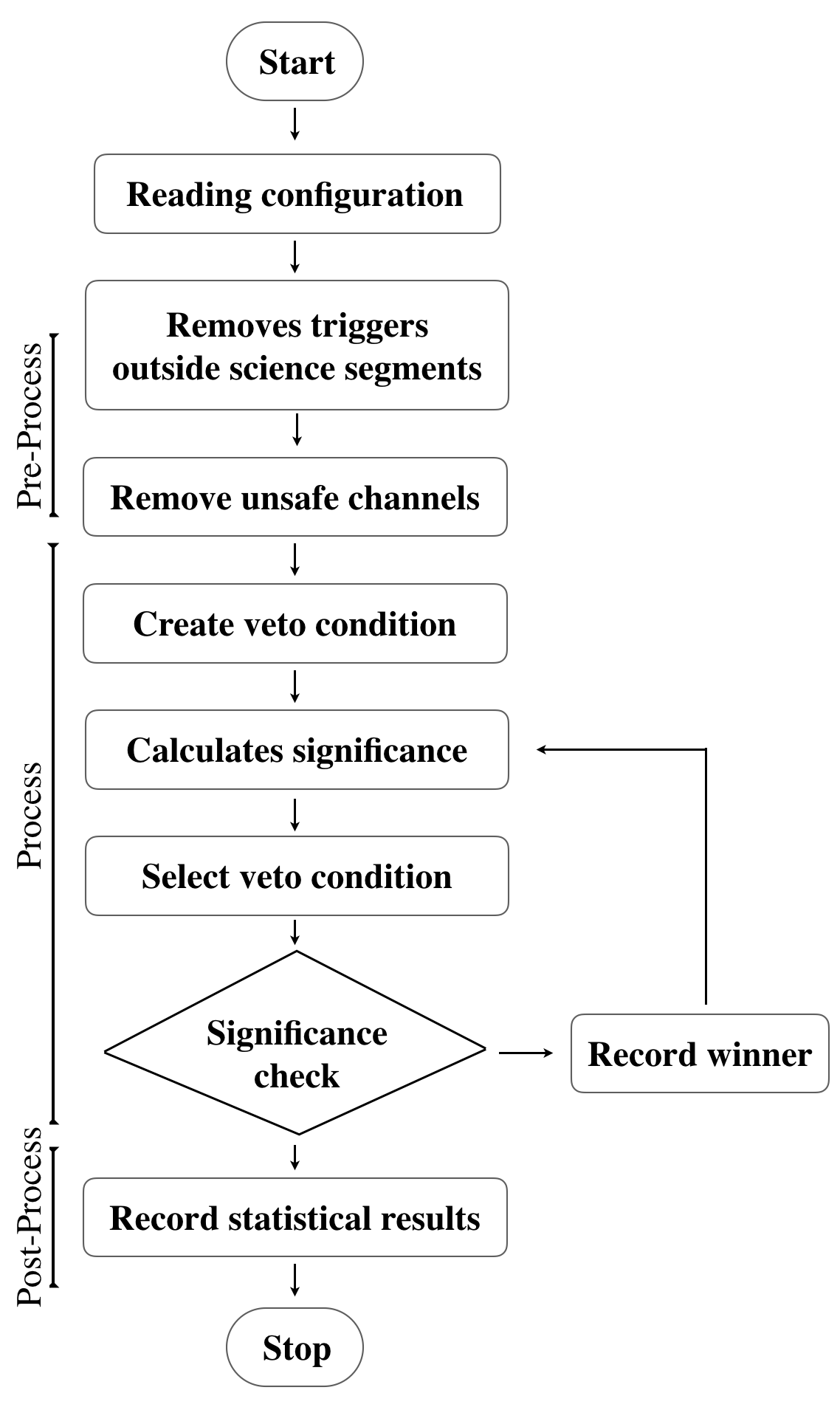}
\caption{Flowchart of the h-veto algorithm\cite{d16}.} \label{Fig.5}
\end{figure}

The h-veto method is veto algorithm with the most characteristic features. It produces some potential correlations or mechanisms which are revealed by the hierarchical process, applying the statistical interpretation between two channels. One easily finds that h-veto consists of three steps of main algorithms; \textit{pre-process}, \textit{process} and \textit{post-process}. (Fig.~\ref{Fig.5})

The pre-process is the routine of confirming basic condition and necessary information for analysis such as the time interval, frequency range, SNR threshold of triggers, time window, and significance threshold. It is possible to prevent the gravitational-wave signal from being mistaken for noise by additionally setting the unsafe safe channel list. The safe channel is an auxiliary channel which is basically independent of the main channel. If an auxiliary channel is highly correlated, the channel is classified as unsafe channel and excluded from the actual analysis. While the unsafe channels are excluded from the analysis, h-veto enter the process stage with the conditions given at the initialization.

The process step is the central routine of h-veto. This process is performed step by step over a round, as its name suggests. First, it creates a possible veto condition for all auxiliary channels according to the given time windows and SNR threshold. Based on the generated conditions, the \textit{significances} of the channels are calculated. 

The significance is defined as 
\begin{equation}
S=-\log _{ 10 }{ \left[ \sum _{ k=n }^{ \infty  }{ P(\mu ,k) }  \right]  },
\end{equation}
where $ n $ represents the number of matches in the trigger on the main channel and triggers in the auxiliary channel for a given time ${T}_{tot}$. It is nothing but a statistical indicator that represents the probability of observing as many or more coincident triggers between the auxiliary channels and the main channel with the Poisson distribution
(Fig.~\ref{Fig.4}). ${P}(\mu, k)$ is the Poisson probability distribution function,
\begin{equation}
P(\mu ,k)=\frac { { \mu  }^{ k }{ e }^{ -\mu  } }{ k! } ,
\end{equation}
where $\mu$ represents the expectation value that any trigger of both channels coincide with each other, and it is defined as follows,
\begin{equation}
\mu =\frac { { N }_{ h }{ N }_{ aux }{ T }_{ win } }{ { T }_{ tot } } ,
\end{equation}
where ${N}_{h}$ and ${N}_{aux}$ are the number of triggers for the main channel and auxiliary channel, respectively.

Next, compare the values of all channels and declare the channel with the highest significance as round winner in the given window and SNR range. If the significance value of this round winner is higher than the significance threshold, it proceeds to the next round.

In the second round, it calculates again the significance from the analysis time excluding the time segments corresponding to the coincident triggers of the first round winner channel. This is the core of the h-veto algorithm which analyzes the change of significance and the change of characteristics of noise mechanism by excluding first round winner. If this significance value is lower than the significance threshold, h-veto is interrupted, otherwise the third round starts. As the result, the channel with the highest significance value declares the second round winner. 
After several rounds, if the significance is lower than the criterion, the process step ends and record the statistical information and veto segments.

\begin{figure}[t!]  
\includegraphics[width=6.6cm]{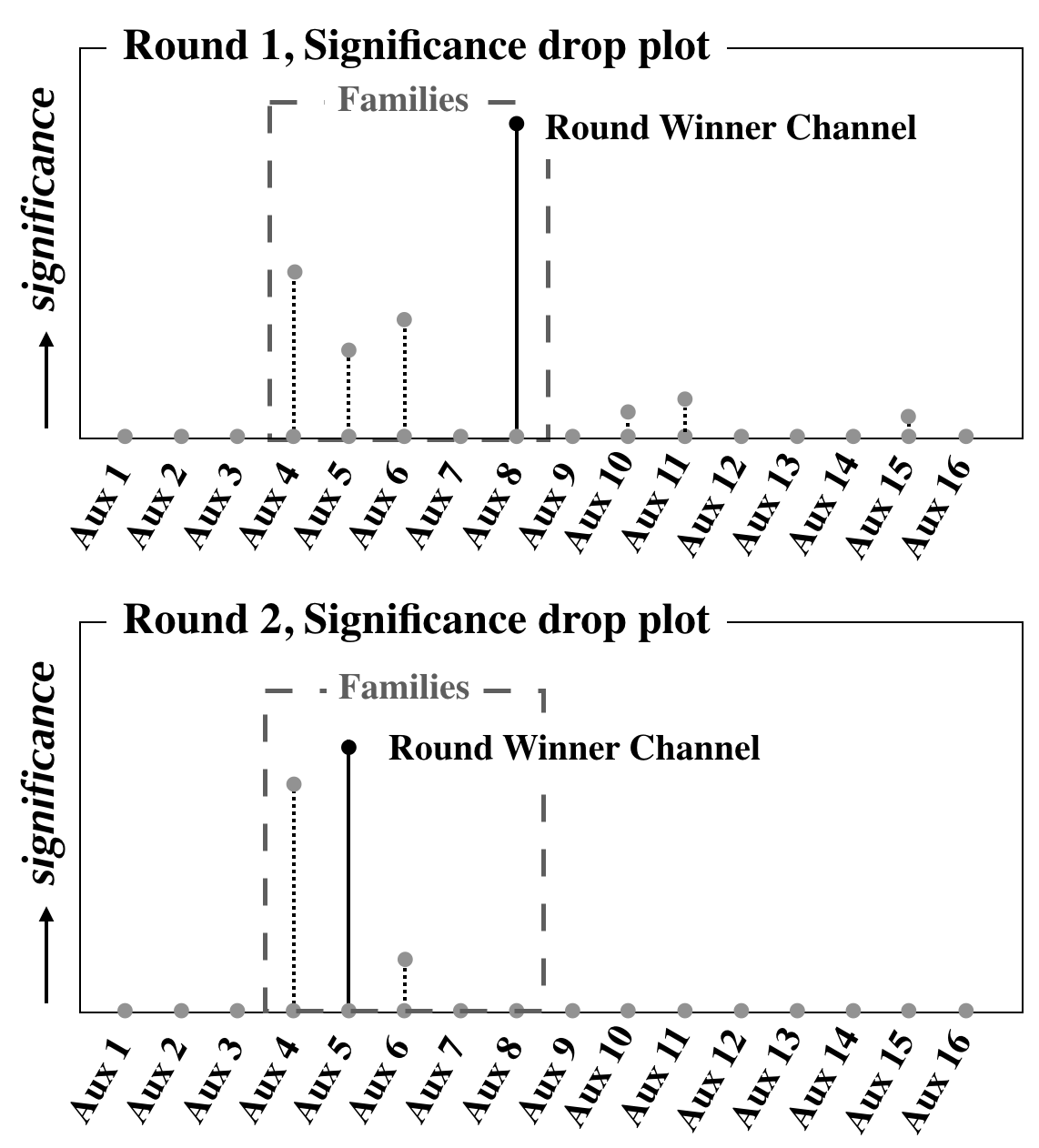}
\caption{Significance drop plots of the h-veto algorithm result\cite{d16}.} \label{Fig.6}
\end{figure}

The final output of the statistical information by h-veto is divided into three types. The first one is the significance drop plot which is the most specific information of this method, the second one is the SNR plot of the vetoed triggers through each round, and the last one is the \emph{deadtime-efficiency} plot to examine the validity of the h-veto results.

A significance drop plot provided at each round shows how the significances of the remaining channels change since the triggers in $h(t)$ correponding to the coincident triggers of the round winner channel are rejected. If some channels exit that changes with the significance of the vetoed channel, these channels are called \textit{families}. If these families channels founded, it implies that the signals in the acquired data at the analysis time are influencing each other on these channels or are highly correlated with each other (Fig.~\ref{Fig.6}). 
A time-trigger SNR plot can be used to see how much of the triggers are correlated with each other and vetoed in each round. 

The deadtime-efficiency plot provides important informations to determine the validity of the veto results delivered by h-veto. Deadtime is the fraction of vetoed time for the whole time, $\rm{DeadTime}=(100\times {T}_{vetoed})/{T}_{tot}$. Efficiency is the fraction of the vetoed trigger for the full triggers in the main channel, $\rm{Efficiecy}=(100\times {N}^{GW}_{vetoed})/{N}^{GW}_{tot}$. If the ratio of two values is greater than 1, then this veto is useful, and if it is close to 1, it no different from vetoing the trigger arbitrarily as follows,

\begin{eqnarray}
&&{\rm if}~~ \frac{\rm Efficiency}{\rm DeadTime} \gg 1 ,~~{\rm veto~is~useful}, \nonumber\\
&&{\rm if}~~ \frac{\rm Efficiency}{\rm DeadTime} \sim  1,~{\rm same~time~removed~at~random}. \nonumber
\end{eqnarray}

	\subsubsection{Used Percentage Veto method}

The UPV is a method of presenting and analyzing how closely the triggers coincide between the main channel and the auxiliary channels. This determines the time-coincidence trigger that satisfies the $\pm 1$ coincidence window that increases the trigger SNR threshold from 50 to 5\,000 sequentially. The final outputs of the results are \textit{Used Percentage}, \textit{Efficiency-Deadtime ratio} and \textit{Random Used Percentage} indices.

Used Percentage is an indicator of the rate at which triggers match between the main channel and the auxiliary channels defined as follows. 
\begin{equation}
{\rm Used~Percentage}(\rho )\equiv 100\times \frac { { N }_{ coinc }^{ aux }(\rho ) }{ { N }_{ tot }^{ aux }(\rho ) },\nonumber 
\end{equation}
where ${ N }_{ coinc }^{ aux }(\rho )$ is the number of triggers in the auxiliary channel and main channel at the trigger SNR threshold $\rho$, and ${ N }_{ tot }^{ aux }(\rho )$ is the total number of triggers in the auxiliary channel. This value reveals how the two channels are correlated.

Deadtime-Efficiency ratio is the value used to determine the effectiveness of the veto with the threshold as already noticed in the previous subsection for h-veto process. UPV also uses this indicator. 
Random Used Percentage is a value related to determining how adequately the Used Percentage calculated above is manifesting the correlation between the two channels. This describes the Used Percentage value when the triggers of the auxiliary channel randomly distributed,
\begin{equation}
{\rm\small Random~Used~Percentage}(\rho )\equiv {\rm Deadtime}(\rho )\times \frac { { N }_{ tot }^{ GW } }{ { N }_{ tot }^{ aux }(\rho ) }.\nonumber
\end{equation}

Once these three indicators are calculated for each threshold, the UPV creates a veto segment for each channel based on the criterion. This describes the time window around the trigger peak time using a trigger SNR threshold with Used Percentage greater than 50 \% for each the auxiliary channel and is rounded to the integer time according to the convention.

\subsection{Channel Safety Study}

When auxiliary channels are used for sensing and vetoing noise signals in $h(t)$ data, 
the auxiliary channels are guaranteed not to have physical couplings or statistically strong correlations to the main degrees of freedom for $h(t)$ calibration system including $h(t)$ channel itself.
If the gravitational-wave passes through the arms of the interferometer, the photon detector of the Michelson interferometer observes the variation of the position of the mirror at the end of the arm directly. At this time, if the auxiliary channel that behaves the main channel in the same way is classified as a safe channel for the veto process, the gravitational-wave signal acquired from the main channel is considered to be noise and is likely to be vetoed. Therefore, it is important to evaluate the safety of channels to prevent this mischance.
%
%


In order to guarantee not to systematically veto a potential GW signal, hardware~injections are used for testing the safety of the auxiliary channels to be used in veto process. Hardware~injections are simulated signals which mimic expected GW signals injected into $h(t)$ data by moving interferometer mirrors to induce the differential arm length \cite{brown2004}. 
During the analysis time when the hardware injections are injected, if the number of triggers rejected by veto segments generated by a given auxiliary channel is greater than expected by chance, the auxiliary channel is considered unsafe. Since the unsafe channel has non-negligible sensitivity to hardware injections as well as a expected gravitational wave signal, it needs to conduct further follow-up study on the veto segments and the auxiliary channel. 

For example, in h-veto case\cite{d16}, if the statistical significance of the coincidences between hardware injection times and auxiliary channel triggers is greater than 3, it is considered to have a statistically meaningful high correlation which increases the possibility to be an unsafe channel. The coherence analysis is used to determine the unsafe channel with the statistics presented by the significance drop plot, the deadtime-efficiency plot, and the time-triggered SNR plot. The estimated channel is finalized through the data quality evaluation process.




\section{Summary}
\label{sec:6}
We have provided a pedagogical review on gravitational-wave and the related astrophysical sources, gravitational-wave detector including advanced technologies, and status of current observations. We also reviewed the gravitational-wave data analysis pipelines and detector characterization.
It is expected that the upcoming observing runs with aLIGO, AdV, and KAGRA reveal the mystery of the Universe by searching various gravitational-wave events such as neutron star-black hole (NSBH) binary or core collapse SNe. Furthermore we will understand clearly the unanswered questions about internal structure of neutron stars or physics of black holes with the observation of gravitational-wave. Together with enhancing gravitational-wave telescopes, the associated techniques for data analysis and detector characterization will be improved.
The truth is out there and the gravitational-wave astronomy era has just began.


\begin{acknowledgments}
J.J.Oh would like to thank Jessica McIver, Reed Essick, and Stefan Hild for fruitful comments and advices on this article. J.J.Oh would like to thank APCTP to elevate the gravitational-wave physics and astronomy by supporting the summer school as a part of APCTP academic program. P.-J.Jung would like to thank Kajita Takaaki, Kazuhiro Hayama, and Nobuyuki Kanda for the opportunity and hospitality during his visit to KAGRA.  The work of K.-Y. Kim and P.-J. Jung was supported by Basic Science Research Program through the National Research Foundation of Korea(NRF) funded by the Ministry of Science, ICT $\&$ Future Planning(NRF- 2017R1A2B4004810) and GIST Research Institute(GRI) grant funded by the GIST in 2018. YMK was supported by NRF grant funded by the 
Korea government (MSIP) (No. 2016R1A5A1013277).
\end{acknowledgments}






\end{document}